\documentstyle{camk}

\newif\ifAMStwofonts
\AMStwofontstrue
 
\ifoldfss
  \ifCUPmtlplainloaded \else
    \NewTextAlphabet{textbfit} {cmbxti10} {}
    \NewTextAlphabet{textbfss} {cmssbx10} {}
    \NewMathAlphabet{mathbfit} {cmbxti10} {} 
    \NewMathAlphabet{mathbfss} {cmssbx10} {} 
  \fi
  \ifAMStwofonts
    \ifCUPmtlplainloaded \else
      \NewSymbolFont{upmath} {eurm10}
      \NewSymbolFont{AMSa} {msam10}
      \NewMathSymbol{\upi}     {0}{upmath}{19}
      \NewMathSymbol{\umu}     {0}{upmath}{16}
      \NewMathSymbol{\upartial}{0}{upmath}{40}
      \NewMathSymbol{\leqslant}{3}{AMSa}{36}
      \NewMathSymbol{\geqslant}{3}{AMSa}{3E}

       \let\le=\leqslant
       
    \fi
  \fi
\fi 
 
\ifnfssone
  \newmathalphabet{\mathit}
  \addtoversion{normal}{\mahit}{cmr}{m}{it}
  \addtoversion{bold}{\mathit}{cmr}{bx}{it}
  \newmathalphabet{\mathbfit} 
  \addtoversion{normal}{\mathbfit}{cmr}{bx}{it} 
  \addtoversion{bold}{\mathbfit}{cmr}{bx}{it}
  \newmathalphabet{\mathbfss} 
  \addtoversion{normal}{\mathbfss}{cmss}{bx}{n}
  \addtoversion{bold}{\mathbfss}{cmss}{bx}{n}
  \ifAMStwofonts
    \ifCUPmtlplainloaded \else
      \UseAMStwoboldmath
      \makeatletter
      \new@mathgroup\upmath@group
      \define@mathgroup\mv@normal\upmath@group{eur}{m}{n}
      \define@mathgroup\mv@bold\upmath@group{eur}{b}{n}
      \edef\UPM{\hexnumber\upmath@group}
      \new@mathgroup\amsa@group
      \define@mathgroup\mv@normal\amsa@group{msa}{m}{n}
      \define@mathgroup\mv@bold\amsa@group{msa}{m}{n}
      \edef\AMSa{\hexnumber\amsa@group}  
      \makeatother
      \mathchardef\upi="0\UPM19
      \mathchardef\umu="0\UPM16
      \mathchardef\upartial="0\UPM40
      \mathchardef\leqslant="3\AMSa36
      \mathchardef\geqslant="3\AMSa3E

       \let\le=\leqslant

    \fi
  \fi
\fi 
   
\ifnfsstwo
  \DeclareMathAlphabet{\mathbfit}{OT1}{cmr}{bx}{it}
  \SetMathAlphabet\mathbfit{bold}{OT1}{cmr}{bx}{it}
  \DeclareMathAlphabet{\mathbfss}{OT1}{cmss}{bx}{n}
  \SetMathAlphabet\mathbfss{bold}{OT1}{cmss}{bx}{n}
  \ifAMStwofonts
    \ifCUPmtlplainloaded \else
      \DeclareSymbolFont{UPM}{U}{eur}{m}{n}
      \SetSymbolFont{UPM}{bold}{U}{eur}{b}{n}
      \DeclareSymbolFont{AMSa}{U}{msa}{m}{n}
      \DeclareMathSymbol{\upi}{0}{UPM}{"19}
      \DeclareMathSymbol{\umu}{0}{UPM}{"16}
      \DeclareMathSymbol{\upartial}{0}{UPM}{"40}
      \DeclareMathSymbol{\leqslant}{3}{AMSa}{"36}
      \DeclareMathSymbol{\geqslant}{3}{AMSa}{"3E}

       \let\le=\leqslant

    \fi
  \fi  
\fi 
      
\ifCUPmtlplainloaded \else
  \ifAMStwofonts \else 
    \def\upi{\pi}
    \def\umu{\mu}
    \def\upartial{\partial}
  \fi
\fi
 
\title{Implication of the discovery of the proper motion of the optical
counterpart of  GRB 970228 for the models of 
gamma-ray bursts}
 
\author[B. Czerny, T. Bulik, M. Sikora, O. Vilhu]
       {B. Czerny$^1$, T. Bulik$^{1,2}$, M. Sikora$^1$ and O. Vilhu$^3$\\
        $^1$ Nicolaus Copernicus Astronomical Center, Bartycka 18, 00-716
        Warsaw, Poland\\
        $^2$ Department of Astronomy and Astrophysics, 
        University of Chicago, 5640 South Ellis Ave, Chicago IL 60637\\        
        $^3$ Observatory, University of Helsinki, Box 14, 00014 Helsinki, 
        Finland\\}
\shorttitle{Galactic nature  of GRB 970228 ???}
 
\begin{document}
 
\maketitle
 
\begin{abstract} 
The accurate position determination of GRB~970228 by the Beppo/SAX
Satellite led to the discovery of a fading  X-ray and optical
counterparts of the burst. About a month  after the GRB event the
proper motion of the ejecta was detected and  the extended optical
source has faded below the Keck detection level. We analyze these
observations in the framework  of the most popular gamma ray burst
models. We find a lower limit on
the distance to GRB~970228 $d>6.6$~pc.
We estimate the amount of energy required to explain the 
motion of a point source and the variability of the extended source.
We find that the cosmological models are ruled out.
The Galactic corona models suffer difficulties if the 
distance to the burst is $> 100$~kpc or the variability
of the extended source is directly connected with the burst
No  constraints are found for GRBs originating between 
$6.6$~pc and $\sim 10$~kpc. 
\end{abstract}
 
\begin{keywords}
gamma-rays: bursts- X-rays: transients - optical: 
transients - stars:neutron
\end{keywords}
 
\section{Introduction}
 
Almost three decades after their discovery \cite{kso:73} the origin
of the gamma ray bursts (GRBs) is still under debate  
\cite{Pacz-deb,Lamb-deb}. The cosmological distance scale was
strongly suggested by the high degree of isotropy of GRB
distribution measured by BATSE, as well as  by the roll-over in the
brightness distribution \cite{batse3b}. It has been shown that both
these features can also be reproduced by galactic corona models
invoking high velocity neutron stars \cite{bl:95a,bl:95b,LiDT:93}.
Some authors also entertain the idea of GRBs originating from the
Oort cloud \cite{Oort}.

The break-through in this field came with the detection  of the
burst GRB 970228 on February 28, 1997  by a new Italian-Dutch
satellite Beppo/SAX  \cite{cffzn:97}. A fast and accurate
measurement of the source position with the X-ray camera yielded 
an error box  only $\sim 2$ arcmin across \cite{cfpcf:97}. This has
been followed by numerous optical observations, and a detection of
an optical variable \cite{ggpms:97}. Next came a report of a
discovery of  an extended source \cite{ggpst:97} at the position of
the optical transient, suggesting a distant galaxy as a source  
of the GRB event \cite{Paradijs:97}. This discovery was soon
confirmed by the HST observation on March 26 \cite{slpm:97},  which
has shown the existence of a point source and an extended
component.  A subsequent HST observation has been performed on April
7  \cite{slpmp:97}.  An attempt to obtain  the redshift of the
extended object  using Keck telescope yielded a spectrum with no
identifiable lines \cite{thc:97}. Analysis of the light curves in
different wave bands show that emission follows $\approx t^{-1}$
\cite{Reichart:97} which has been found to be consistent with a
relativistic fireball afterglow \cite{WMR:97}.
 
A comparison of the two Hubble observations has shown that  the
point object has a significant proper motion of $\approx 540
$~mas~year$^{-1}$ \cite{iue6629}. Furthermore, it has been 
reported \cite{iue6631} that the extended source has faded  below
detection limit in the R band  using the Keck telescope on April 5
and 6, indicating also a variability on a $0.5''$ scale, and
undermining the identification of the extended object with the 
host galaxy.

In this paper we discuss the direct  implications of these
observations  on the gamma-ray burst distance scale and their models.
In section 2 we derive a lower limit on the  distance to GRB~970228
from the parallax and define distances for which relativistic effects
are important, in section 3 we discuss the galactic corona
models. In section 4  we overview the requirements for the
cosmological model,  and finally we discuss the results in
section 5.
 
\section{Limits on the distance scale}

\subsection{Minimum distance to GRB 970228}
 
The possibility that gamma ray burst are very local phenomenon is
not frequently discussed since there is no specific suggestion of
how the gamma rays may actually be produced in the result of, for
example, the cometary collisions. Also the present knowledge about
the structure of the Oort cloud does not seem to be in agreement
with the statistical properties of gamma bursts \cite{Maoz:93}.
However, the conclusion  of that study was not firm  so in any case
it is interesting to consider  the lower limit for the distance to
the source which can be derived from the discovered proper motion.
 
If the source is really as close to the Earth as the Oort cloud its
apparent motion consists of two components: (i) proper motion and
(ii) the effect of parallax due to the orbital motion of Earth. The
second effect is unavoidable and it is difficult to suspect that
the two independent movements might simply cancel.
 
The source is located at R.A. = 5h01m44s, Decl.=+11$^\circ$46'.7. 
Therefore the angle $\theta_{V-GRB}$ between the orbital motion of the
Earth  and the direction to GRB 970228 is only
$\theta_{V-GRB}\approx 23.8^\circ$ on March~26. 
The component of the Earth velocity
perpendicular to the direction towards the source is equal
$V_{Earth}  \sin\theta$. 
For an observation lasting 13 days we assume that the Earth
moves along a straight line.
The expected parallactic motion expressed in
''year$^{-1}$  for a source at a distance $D$ measured in pc is given
by 
\begin{equation}
 \pi = {V_{Earth} t \sin\theta_{V-GRB} \over D} = {2.5 \over D}
\end{equation}
The direction of the observed proper motion has been reported  as
``south-east'' while the expected direction of parallactic motion is almost
exactly east at this time of year. Therefore  only a fraction  equal
$0.38$''year$^{-1}$ out of the total velocity of  $0.54$''year$^{-1}$  can be 
attributed to the paralax. Taking $\pi = 0.38$, as observed, we obtain the
minimum distance  to the source 
\begin{equation} 
 D_{min}= 6.6 \,{\rm pc}.
\end{equation}

In principle the proper motion of the object may compensate for the
orbital motion of the Earth and the distance to the object can be
as small as for example $0.1$~pc. However, it requires that the peculiar
velocity of GRB~970228 be equal to that of the Earth with the
accuracy of $\sim $ 2 \%. Although the amplitude of the velocity
($\sim 30$ km\,s$^{-1}$) is of the same order as the dispersion velocity in
the Sun vicinity the precise coincidence of the direction with the
orbital motion of the Earth is improbable.
 
The minimum value of $6.6$~pc for the distance of GRB~970228 is much
stronger  limit that the lower limit of $12000$~A.U obtained from
triangulation \cite{Hurley:97}. 
It clearly excludes the Oort cloud as the birth
site of the gamma bursts.

\subsection{Characteristic  scales}

The observed variability of the observed counterparts:
the moving point source and the fading extended object
introduces characteristic regimes in the distance scales.
The proper motion of the point source on the sky is
$0.54$''year$^{-1}$. The inferred linear velocity perpendicular
to the line of sight is
\begin{equation}
v_\perp^{ps} \approx  10^{-5} {D\over 1{\rm pc}} c \, .
\end{equation}
Thus the distance at which the apparent velocity of the point
source on the sky reaches the speed of light is $D_1 = 100$~kpc.
The angular size of the extended source is $\approx 0.5$'', and 
we take a conservative assumption that it varies on the timescale of 
a month. A typical apparent linear velocity i.e. the ratio of the
linear size to the observed variability scale is
\begin{equation}
v_\perp^{ex} \approx 10^{-4} {D\over 1{\rm pc}} c \, .
\end{equation}
This define a distance at which $v_\perp^{ex}=c$, so $D_2 = 10$~kpc.

We can now consider three regions in the space of possible distance to
GRB~970228: A. the nonrelativstic region $D < D_2$, B. the mildly
relativistic region $D_2 < D < D_1$ and C. the relativistic one $D_1 <
D$. It should be noted that regions B and C are  not automatically
excluded since  apparent superluminal motion can be observed due
to relativistic beaming. The apparent motion of the ejecta as projected on
the sky
proceeds with the velocity 
\begin{equation} v_{\perp} = {c \beta 
\over 1 - \beta \cos \theta} \sin\theta  \approx c \Gamma\, ,
\end{equation} 
where $\theta $, the angle between the ejecta axis
and the line of sight, is taken $\sim 1/\Gamma$.

\section{Galactic halo distances}

In the framework of these models  GRBs originate from high velocity 
neutron stars, with characteristic velocities $\sim 1000 $~km~s$^{-1}$.
Such objects would leave the  Galactic plane (the escape velocity from the
Galaxy  is $\sim 600 $km~s$^{-1}$) and form a Galactic corona  
\cite{Lamb-deb,LiDer:92}.  Observations of isotropy of gamma-ray burst
distribution on the sky \cite{batse3b} provide a strong constraint on the
distance to a typical burst. This constraint can be consistent only with
models characterized by strong source evolution (also called the delayed
turn-on, furthermore DT)  or within the beaming model \cite{LiDer:92} 
where the bursting direction is aligned with the direction of the initial
kick velocity, furthermore BM. The isotropy constrains the beaming angle to
be $15^\circ <\theta_b <25^\circ$ \cite{bl:95b,DL:97}. The estimates of the
distance to the farthest burst seen by BATSE varies from $80$~kpc to
$350$~kpc \cite{DL:97,bl:95b}. Physical models of gamma ray bursts in the
Galactic corona invoke magnetic instabilities in super magnetic neutron
stars \cite{DT:92,PRR:95,LBC-SD}, or accretion of a planetesimal on a
neutron star \cite{ColgateL:95}. Within the framework of each of the
scenarios a blast wave or a jet can follow a gamma ray burst event.

Since GRB~970228 was a relatively bright burst we  assume that it is
located at the distance of 100 kpc or less, i.e. in the mildly
relativistic region.   This estimate of $\Gamma$ depends, of course,
on the assumed distance and varies from $\approx 1.1$ for the distance
of 50 kpc, $1.9$ for $100$~kpc to $\approx 3$ for 300 kpc. The
variability of the extended source requires  Lorentz $\Gamma$ factors 
in the range from $5$ to $30$.

Given the above calculated value of $\Gamma$ we can estimate that  the
extent of the visible emission of the jet comes from  the angle smaller
than  $\Gamma^{-1} \approx 30^\circ$. Assuming that  the visible jet is
directly associated  with the gamma-ray burst we immediately obtain that if
the expansion is isotropic we only see a fraction $f =0.25
\Gamma^{-2}\approx 0.06$ of the emitting surface.  However, since the
source is moving we know that the expansion is not spherical.  In the
framework of the DT model this places an interesting limit, requiring that
there is $4\Gamma^2 \approx 16$ times more sources of the galactic corona
GRBs  or that they burst $16$ times more often.  Such a limit does {\em
not} arise within the framework of the BM model proposed by  \cite{DT:92}
since the direction of emission is not randomly oriented. It has to be also
noted that we find a tantalizing agreement of two numbers: the estimated
angular width of the jet ($30^\circ$) and the beaming angle  required by
the isotropy of the BATSE bursts ($15^\circ - 25^\circ$) in the framework
of this model.

The optical transient has been observed for $\Delta t_{obs} \approx
1$~month. In this time the ejecta must have traveled a distance of
\begin{equation}
r \approx  c \Delta t_{obs} \Gamma^2 
           \approx  10^{17} \Gamma^2 \, {\rm cm} \, .
\end{equation}
 A relativistic shock collects matter which slows it 
down while ploughing through  space. The amount of matter 
collected by the ejecta can be estimated as
\begin{equation}
\Delta M = {\pi\over 3} a^2 r\rho \approx a^2 r \rho\, ,
\end{equation}
where $a = \Gamma^{-1}  r$ is the perpendicular size of the
cone sweeping by the ejecta.
 Thus we obtain $\Delta M = 10^{51} \Gamma^4 
 (\rho/1{\rm g\, cm}^{-3})$~g.
 A relativistic shock will  be efficiently decelerated if
it collides with a mass $\Delta M > m_e \Gamma^{-1}$
where $m_e$ is the mass of the ejecta.
Thus we obtain a constraint
on  the mass of the ejecta 
\begin{equation}
m_e> \Delta M \Gamma =
10^{51} \Gamma^5  (\rho/1{\rm g\, cm}^{-3}){\rm ~g}\, .
\end{equation}
 The energetic requirement for a burst source to accelerate 
 $m_e$ to a Lorentz factor $\Gamma$ is 
 \begin{equation}
 E \approx m_e c^2 \Gamma >
10^{42} \Gamma^6  {\rho\over 10^{-30}{\rm g\, cm}^{-3} }\, {\rm erg}\,
 ,
 \label{energy}
 \end{equation}
where we scaled the  density to a typical 
value  for the intergalactic 
space. 
 
For the expected range of $1< \Gamma< 3$ for the  motion of the point
source these models are on the  edge of viability  with energetic
requirement varying from  a comfortable $10^{42}$~ergs to rather
constraining $10^{45}$~ergs. The variability of the extended source 
increases these requirements by a factor of at least $10^4$ and up to
$10^6$. Note that the above estimates are very conservative  since the
density in the Galactic halo can be higher.

\section{Extragalactic origin}
 
Cosmological  origin of GRB explains in a natural way both the
impressive isotropy of the burst distribution as well as the
roll-over of the brightness  distribution  for low fluxes which
leads to  $V/V_{max}\approx 0.3$ instead of $0.5$ expected for a
homogeneous distribution in Euclidean space \cite{batse3b}.
 
Within this frame, the most popular family of models are models
based on sudden merger of two neutron stars or a neutron star  and
a black hole \cite{Pacz-Acta}. Such a merger is claimed to lead 
to  ejection of a fraction of the disrupted star at
ultra-relativistic speeds corresponding to the bulk Lorentz
factor, $\Gamma > 100$ \cite{MR-blast}.  The ejecta are usually
considered to be quasi-spherical or mildly collimated into
direction of the angular momentum vector of the system. We consider
below the implication of the detected proper motion of the optical
point source to this general scenario.
 
Identifying  the proper motion of the optical source as a motion of
relativistic ejecta observed at very small angle to the ejecta
axis,  we obtain that the  opening angle of the ejecta must be, at
most, of the order  of the Doppler beam ($\le 1/\Gamma$)
(otherwise,   one would see on the sky the expanding source, rather
than the propagating  source). Let us assume that the distance to
the GRB 970228 is equal to $1$~Gpc which corresponds to a rather
moderate cosmological  redshift of $\sim 0.16$. It is a very
conservative assumption since on the basis of the gamma-ray
brightness the source may be possibly placed at $z \approx 0.3$
\cite{WMR:97}, provided it is a standard cosmological candle. Even
larger distance is suggested by weakness or  absence  of the host
galaxy.  Thus, for  a
distance of 1 Gpc measured by HST the proper motion  $540$''year$^{-1}$
translates  into the bulk Lorentz factor  $\Gamma \sim 8500$.

For such large Lorentz factor the 
ejecta travel for a very large distance
$r > 10^{25} (\Gamma/10^4)^2 {\rm cm}$.
Thus the ejecta can not avoid interacting with 
the matter swept over such a journey and
the mass of the gas collected 
is at least $\Delta M \approx 10^4 (\Gamma/10^4)^4$~M$_\odot$.
The energetic requirements of equation~\ref{energy} 
can be applied  to the case of cosmological models as well.
In particular when $\Gamma = 10^4$, we find that
\begin{equation}
E > 10^{65} (\Gamma/10^4)^6 (\rho/10^{-30}) \, {\rm erg}.
\end{equation}
Such a requirement seems to rule out all of the cosmological models.
The variability of the extended source pushes the above requirement
up by six orders of magnitude.

\section{Discussion and conclusions}
 
%
%
%
%
%
%

The proper motion of the discovered point-like optical counterpart imposes
severe constraints on the origin of gamma ray bursts. In the following
discussion we make an assumption that the moving point source is directly
related to the GRB. The relation of fading extended object to the GRB event
seems less clear. Thus we discuss various possibilities, assuming that
either it is or it is not directly related to the  GRB.

The comparison of the observed proper motion with the parallax expected at a
given distance puts a firm limit on the minimum distance to the source
equal $6.6$~pc. This conclusion rules out all models  connecting the gamma
burst phenomenon with Oort cloud.

The assumption of cosmological distances to gamma ray bursts leads also to
serious problems. Observed proper motion requires huge Doppler factors
($\Gamma \sim 10^4$) for the bulk motion of the radiating gas. The time
evolution of the accompanying extended source require the Doppler factor
even by a factor of 10 higher. We have shown that the energy required to 
propagate such a highly relativistic gas is eleven orders of magnitude
higher than that available in a collision of two neutron stars.

Also the models locating gamma ray bursts in the Galactic halo  may have
difficulties. The energetic requirements can be satisfied if the typical
distance to a GRB is less than $100$~kpc. Such a distance scale is
marginally consistent with the BATSE observations of isotropy of the GRB
distribution. The delayed turn-on models (DT) require that the sources
burst $4\Gamma^2\approx 16$ more times, while in the beaming model (BM) no
such requirement arises.

The $\Gamma$ factors required by the variability of the extended source
make the energetic requirements prohibitive for the galactic halo models of
GRBs.

The apparent motion of the point source or the variability  of the extended
source do not constrain the models for distances in the nonrelativistic 
region A., i.e $6.6~{\rm pc} < D < 10$~kpc. For example we cannot  rule out
the moderately local origin of the gamma bursts. If the bursts come from
the distances of order of 5--100~pc and the roll-over of the brightness
distribution is an apparent effect caused by the detection threshold, like
suggested by  \cite{bk:95}, who argues that  such a geometrical
distribution may also be acceptable at the present stage.

All these conclusions are based on assumptions that (i) the  identification
of the optical counterpart of the gamma ray burst was correct and (ii) the
measurement of the proper motion of the point-like optical counterpart was
undoubtful. Unfortunately, further observations of GRB 970228 are difficult
since the source approaches the Sun on the sky. This part of the sky will
be visible again in a few months but the source is rapidly fading at all
frequencies so its future detection may be impossible.

Therefore the confirmation of our conclusions would rather come from the
detection of similar behavior of the optical counterparts of other gamma
bursts well localized by Beppo/SAX. Since several such bursts a year are
expected at least one or two should be bright enough to allow the accurate
measurements and similar analysis.

\section*{Acknowledgements}
 
We would like to thank the Danish Space Research Institute where the idea
of this paper has been born in a discussion over lunch. This work was
supported in part  by grants 2P03D00410 (BCZ), 2P02D00911 (TB), and
2P03D01209 (MS) of the Polish State Committee for  Scientific
Research.

\ \\
This paper has been processed by the authors using the Blackwell
Scientific Publications \LaTeX\  style file.
 
\end{document}